\documentstyle[12pt,openbib]{article}
\begin{document}
%%%%%%%%%%%%%%%%%%%%%%%%%%%%%%%%%%%%%%%%%%%%%%%%%%%%%%
\def\f{\frac}
\def\th{\theta}
\def\de{\delta}
\def\D{\Delta}
\def\p{\partial}
\def\La{\Lambda}
\def\up{\Upsilon}
\def\be{\beta}
\def\al{\alpha}
\def\ka{\kappa}
\def\s{\Sigma}
\def\g{\gamma}
\def\om{\omega}
\def\ep{\epsilon}
\def\va{\varepsilon}
%%%%%%%%%%%%%%%%%%%%%%%%%%%%%%%%%%%%%%%%%%%%%%%%%%%%%%%%%%%%%%%%%%%%%%%%%%%%
\title{\bf Strange Stars with Realistic Quark Vector Interaction and  
Phenomenological Density-dependent Scalar Potential}
%%%%%%%%%%%%%%%%%%%%%%%%%%%%%%%%%%%%%%%%%%%%%%%%%%%%%%%%%%%%%%%%%%%%%%%%%%%%
\author{ Mira Dey$^{1,2,\dagger}$, Ignazio Bombaci$^{3}$, Jishnu
Dey$^{4,\dagger}$,\\           Subharthi Ray$^{1,4,\dagger}$ and B. C.
Samanta$^{5}$ }
%%%%%%%%%%%%%%%%%%%%%%%%%%%%%%%%%%%%%%%%%%%%%%%%%%%%%%%%%%%%%%%%%%%%%%%%%%%%
\vspace{.5 cm} \date{\today } \maketitle

%body of paper
%\newpage
%%%%%%%%%%%%%%%%%%%%%%%%%%%%%%%%%%%%%%%%%%%%%%%%%%%%%%%%%%%%%%%%%%%%%%%%
\begin{abstract}
We derive an equation of state (EOS) for strange matter, starting from an 
interquark potential which {\it(i)} has asymptotic freedom built into it, 
{\it (ii)} shows confinement at zero density ($\rho_B = 0$) and 
deconfinement at high $\rho_B$, and {\it (iii)} gives a stable  
configuration for chargeless, $\beta$--stable quark matter. 
This EOS is then used to calculate the structure of Strange Stars, 
and in particular their mass-radius relation. 
Our present results confirm and reinforce the recent claim\cite{li,b} 
that the compact objects associated with the x-ray pulsar Her X-1, 
and with the x-ray burster 4U~1820-30 are strange stars.    
\end{abstract}
%%%%%%%%%%%%%%%%%%%%%%%%%%%%%%%%%%%%%%%%%%%%%%%%%%%%%%%%%%%%%%%%%%%%%%%%

Keywords: Strange Quark Matter, compact objects, x-ray bursts.   
\vskip .5cm
%%%%%%%%%%%%%%%%%%%%%%%%%%%%%%%%%%%%%%%%%%%%%%%%%%%%%%
{\it $^{1)}$Dept. of Physics, Presidency College, Calcutta 700 073, 
                    India.}
{\it $^{2)}$An associate member of the International Centre for 
                     Theoretical Physics, Trieste, Italy}.
{\it  $^{3)}$Dipartimento di Fisica, Universit\'a di Pisa and 
           I.N.F.N. Sezione di Pisa, via Buonarroti 2, I-56100 Pisa, Italy.}
{\it $^{4)}$Azad Physics Centre, Dept. of Physics, 
                    Maulana Azad College, Calcutta 700 013, India.}
{\it  $^{5)}$Department of Physics, Burdwan
                   University, Burdwan 713104, India.}
{\it  $^{\dagger}$ Work supported in part by DST 
                  grant no. SP/S2/K18/96, Govt. of India, permanent address: 
                  1/10 Prince Golam Md.  Road, Calcutta 700 026, India.}
%%%%%%%%%%%%%%%%%%%%%%%%%%%%%%%%%%%%%%%%%%%%%%%%%%%%%%%%%%%%%%%%%%%%%%%%%%%%
\newpage
%%%%%%%%%%%%%%%%%%%%%%%%%%%%%%%%%%%%%%%%%%%%%%%%%%%%%%
%%%%%%\narrowtext
%%%%%%%%%%\begin{multicols}{2}
%%%%%%%%%%%%%%%%%%%%%%%%%%%%%%%%%%%%%%%%%%%%%%%%%%%%%%%
~\bigskip

 Strange stars (SS) are astrophysical compact objects which are entirely 
made of deconfined {\it u,d,s} quark matter ({\it strange matter}).  
The possible existence of SS is a direct consequence of the conjecture 
\cite{bod,wit}  that strange matter (SM) may be the absolute ground 
state of strongly interacting matter rather than $^{56}$Fe. 
Since this hypothesis was formulated, strange stars have been studied by 
many authors \cite{bc,hzs,afo,dth}, but they remained purely theoretical 
entities. This situation changed in the last few years, thanks to the large 
amount of fresh observational data collected by the new generation of 
x-ray and $\gamma$-ray satellites. In fact, recent studies \cite{li,b} have 
shown that the compact objects associated with the x-ray pulsar Her X-1, 
and with the x-ray burster 4U~1820-30, are good strange star candidates.

    Most of the previous calculations of SS properties used an equation 
of state (EOS) for strange matter based on the MIT bag model. In this 
phenomenological model the basic features of QCD ({\it i.e.} quark 
confinement and asymptotic freedom) are built in. The deconfinement of 
quarks at high density is not obvious in the model.  Some people have 
also expressed more general reservations about the bag model and, in 
particular, about the use of a large value for the bag constant \cite{g}.  
In fact, to get SS configurations with a mass-radius (MR) relation in
agreement with the semiempirical MR relations \cite{li,ht} for the two 
SS candidates mentioned above, one must indeed use large values of the bag
constant $B \simeq 110\;MeV/fm^3$  (which in different units correspond to  
$\bar B \equiv (\hbar c)^3 B  \simeq (170.5\;MeV)^4$, which is still 
small compared with the estimate from QCD sum rules \cite{g,s}.  
But this value is large when compared  with the ``standard value'' 
$B = 56\;MeV/fm^3$ ($(144\;MeV)^4$) which is able to reproduce the 
mass spectrum of light hadrons and  heavy mesons \cite{bag}.  
Large values of B are also not allowed by the requirement SM is stable in 
bulk \cite{fj,mad}: {\it e.g.} for $m_s = 150 MeV$ the permitted values of 
the bag constant are in the range (56 -- 78) $\;MeV/fm^3$ 
(see Fig.1 of ref.\cite{fj}).   However, in general, $B$ has been
considered as an effective parameter, and the bag model should be only
regarded as a simple model with limited connection to QCD. 
%%%%%%%%%%%%%%%%%%%%%%%%%%%%%%%%
\footnote{The point is that QCD sum rules \cite{s} lead to a bag constant,  
which is substantially larger than the value fitted in the simple bag model. 
This theme is stressed, {\it e.g.}, in ref.\cite{g} pag. 367:   
``There do exist a number of modified bag models with a smaller bag 
radius and correspondingly larger B values ~.~.~.~ and the authors claim 
that the sum rule values  would support their model. Such claims are, 
however, not very sensible. In each of these models additional interactions  
are introduced. The energy and pressure associated with these additional 
interactions and that due to the QCD vacuum state are interwoven, such 
that it is not clear which quantity should be compared with the trace 
of the energy momentum tensor." }
%%%%%%%%%%%%%%%%%%%%%%%%%%%%%%%%
We will thus use the bag model, but only to compare our results with it
qualitatively.    

If the compact objects in Her X-1 and 4U~1820-30 are really strange 
stars,  then there will be very deep consequences for both the physics 
of strong interactions and astrophysics. 
The existence of stable nugget of SM would have also profound 
implications in cosmology, in relation to the dark matter
problem, in the sense that many such stars, unlike the two mentioned 
above, may as yet be unobserved. Note however, that by strange stars 
now we mean conglomeration of chargeless beta-stable ({\it u,d,s}) 
quarks held together strongly by gravity, with a surface density about 
five times that of nuclear matter and an interior density which may be 
3 times higher. Witten's original conjecture is still applicable for 
such conglomerates, - that they may remain as relics of the cooling 
of the universe a millionth of a second after its formation \cite{wit}. 
This is largely because chargeless beta-stable ({\it u,d,s})
quarks have a minimum in its EOS at about five times that of nuclear 
matter and can thus form the surface of a SS.

     Motivated by the fundamental importance of this issue, and by the 
criticism to the bag model mentioned above, in the present work we 
investigate the properties of SS, using an EOS for SM based on an 
alternative to the bag model.

Intuitively we know that at high density the quarks should go over 
from their constituent masses to their current masses, thus restoring 
the approximate chiral symmetry of QCD. On the other hand, the interquark 
interaction should be screened in the medium. The latter will give rise 
to deconfinement at high  density.  
To this end we use the following Hamiltonian:
%%%%%%%%%%%%%%%%%%%%%%%%%%%%%%%%
\footnote{In the following we use units where $\hbar = c = 1$} 
%%%%%%%%%%%%%%%%%%%%%%%%%%%%%%%%
\begin {equation} 
H = \sum_i \Big( {\bf \alpha_i \cdot p_i } + \be _i M_i \Big) 
+ \sum_{i<j} \f {\lambda(i) \lambda (j)}{4} V_{ij}
\label{eq:H}
\end{equation}
where we have  two potentials: a scalar and a vector. The scalar 
originates
from the mass term.  The quark mass, $M_i$, is taken to be  density 
dependent
and of the form:
\begin{equation}
M_i =  m_i + (310\;MeV) sech(\nu \f{\rho_B}{\rho _0}), 
\;\;~~~ i = u, d, s.
\label{eq:qm}
\end{equation}
where $\rho_B = (\rho _u+\rho _d+\rho _s)/3$ is the baryon number 
density, 
$\rho _0 =  0.17~fm^{-3}$ is the normal nuclear matter density, and 
$\nu$ is a parameter. 

At high $\rho_B$ the quark mass $M_i$ falls from its constituent value 
to its current one which we take to be \cite{s}:
$m_u =  4 \;MeV,\; m_d = 7 \;MeV,\; m_s = 150 \;MeV$.  
The density dependence introduces a density dependent scalar potential 
and more importantly restores chiral symmetry smoothly at high density 
(Fig.1).

%%%%%%%%%%%%%%%%%%%%%%%%%%%%%%%%%%%%%%%%%%%%%%%%%%%%%%%%%%%%%%%%%%%
%%%%%%%%%%%%%%%%%%    Figure 1
%%%%%%%%%%%%%%%%%%%%%%%%%%%%%%%%%%%%%%%%%%%%%%%%%%%%%%%%%%%%%%%%%%%

It is hard to justify the density dependent mass term except 
qualitatively.
It appears we are the first to suggest such a form. On the other hand at
finite T the restoration of chiral symmetry can be established from 
lattice calculations (see for example \cite{ber}). However, from 
calculations of meson \cite{CVA} or baryon properties \cite{ddt,ddms}, 
it is clear that inside a hadron the quark mass varies with the radius 
even as the density varies with the radius. For example in \cite{CVA}, 
the minimum quark masses are, $m_u\;= \; 186 \; MeV$ and 
$m_s\;= \; 364 \; MeV$ whereas the averaged effective quark masses are 
$m_u\;= \; 305 \; MeV$ and $m_s\;= \; 465 \; MeV$.
It will become clear that in our calculation we are not dealing with 
small variations in the quark density as in a bound-state hadron 
calculation, and thus this density dependence of the scalar potential is  
all the more crucial for us.

  The colour dependent vector potential in  eq.(\ref{eq:H}) is an 
interquark potential originating from gluon exchanges, and the 
$\lambda$-s are the color  SU(3) matrices for the two interacting quarks. 
In the absence of an accurate evaluation of the potential (e.g. from 
large $N_c$ planar diagrams) we borrow it from meson phenomenology, 
namely the Richardson potential \cite{r}. It incorporates the two concepts 
of asymptotic freedom and linear quark confinement. The potential 
reproduces heavy meson spectra and has been used in sophisticated 
calculations involving two-body-Dirac equation derived from Dirac constraint 
mechanism for light as well as heavy mesons \cite{CVA}. 
It has been well tested for baryons in Fock calculations \cite{ddt,ddms}. 
The potential used for the meson and baryon is \cite{r} 
\begin{equation}
V_{ij} =  \f{12 \pi}{27}\f{1}{ln(1 +  {({\bf k}_i- {\bf k}_j)}^2
/\La ^2)}\f{1}{{({\bf k}_i - {\bf k}_j)}^2}  \,\, ,
\label{eq:V}
\end{equation} 
with the scale parameter \cite{s}  $\La = 100 \; MeV$. 
This bare potential in a medium will be screened due to pair creation 
and infrared divergence. The inverse screening length, $D^{-1}$, to the 
lowest order is \cite{baluni}:  
\begin{equation}
(D^{-1})^2 \equiv \f{ 2 \al _0}{\pi} \sum_{i=u,d,s,}k^f_i 
\sqrt{(k^f_i)^2 +
m_i^2}
\label{eq:gm}
\end{equation}
where $k^f_i$, the Fermi momentum of the {\it i-th} quark is obtained 
from the corresponding number density:
\begin{equation}
{k^f_i} = (\rho _i \pi^2)^{1/3}  
\label{eq:kf}
\end{equation}
and $\al _0$ is the perturbative quark gluon coupling. 
To simplify numerical calculations, instead of summing over all the 
individual flavours, we have averaged over the flavours so that
%%%
\begin{equation} (D^{-1})^2 \simeq \f{3\,\,\times \,2 \al _0}{\pi} 
k^f_{av}
\sqrt{(k^f_{av})^2 + m_{av}^2}
\label{eq:gmp}
\end{equation}
%%%
where
\begin{equation}
{k^f_{av}} = (\pi^2 \rho_B  )^{1/3}  = 
         \Big( \f{(k^f_u)^3 +(k^f_d)^3 +(k^f_s)^3}{3} \Big)^{1/3}.
\label{eq:avkf}
\end{equation}   
%%%%%%%%
\begin{equation}
m_{av} = \f{m_u + m_d + m_s}{3} \;\;.
\label{eq:mav}  
\end{equation}
%%%%%%%%%%%%%%
The model we adopt has the following attractive features:
It truly describes deconfined quarks at finite density, through the
Debye screening (DS, in short). At zero density (for an isolated hadron) 

$D^{-1}$ vanishes, leading to confinement.  At finite density, due to 
DS, the gluon polarization acquires a non-zero value leading to 
deconfinement.  
The scalar potential also decreases with density, thus restoring chiral 
symmetry at high density.
The resulting inverse DS  is also plotted  in Fig.1. 

Within the present approach, the energy density of SM can be written 
\begin{equation}
\va =  \va_k +\va _v 
\label{eq:Eden}
\end{equation}
where the kinetic part is given by 
\begin{equation}
\va_k = \f{3}{4\pi^2} \sum _{i=u,d,s}  \Big[ k_i^f\,((k_i^f)^2 +
M_i^2/2)\sqrt{(k_i^f)^2 + M_i^2} - \f{M_i^4}{2}
ln\f{\sqrt{((k_i^f)^2+M_i^2})+k^f_i}{M_i} \Big] 
\label{eq:kin}
\end{equation}
and the potential contribution is given by  
\begin{equation}
\va_v = - \f{1}{2  \pi^ 3}\sum_{i,j} \int_{-1}^{+1} dx
\int_0^{k_j^f}{k_j}^2\int_0^{k_i^f} {k_i}^2f(k_i,k_j,M_i,M_j,x)
V[D^{-1},{({\bf k}_i -{\bf k}_j)}^2]dk_jdk_i
\label{eq:PE}
\end{equation} 
where V is the screened Richardson potential. In other words, 
${({\bf k}_i -{\bf k}_j)}^2$ in  eqn. (\ref{eq:V}) is replaced by     
$ [{({\bf k}_i -{\bf k}_j)}^2\; +\; D^{-2}] $   and 
\begin{equation}
f({k_i},{k_j},M_i,M_j,x) = (e_i.e_j  + 2.k_i.k_j.x +
\f{k_i^2.k_j^2}{e_i.e_j}) \f{1}{(e_i-M_i)(e_j-M_j)}
\label{eq:f}
\end{equation} with
\begin{equation}
e_i  =  \sqrt{k_i^2 + M_i^2} + M_i 
\label{eq:ei}
\end{equation}
Then we calculate the self consistent chemical potentials to satisfy
the $\beta$--equilibrium  and charge neutrality conditions, 
\begin{equation}
\mu _d = \mu _s \, ,   ~~~~~~~~ \mu _d = \mu _u \,+ \mu _e  ~;
\label{eq:sd}
\end{equation}
\begin{equation}
2 (k^f_u)^3 - (k^f_s)^3 - (k^f_d)^3 -  (k^f_e)^3 =  0
\label{eq:ch}
\end{equation}
where $\mu$-s are the chemical potentials of {\it u,d,s} quarks and the
electron, $e$. We assume that the neutrinos have left the system 
($\mu_\nu = 0$). 

If $m_e$ is the electron mass, ${k^f_e}$ is obtained through
\begin{equation}
{k^f_e} = \sqrt{\mu _e^2 -m_e^2}\; .
\label{eq:ekF}
\end{equation} 
To achieve the conditions (\ref{eq:sd},\ref{eq:ch}), we have the 
additional difficulties originating from the density dependence of the 
quark masses and DS length. 

The chemical potential for the {\it i-th} quark is given by :
\begin{equation}
\mu _i = \sqrt{M_i^2 + (k^f_i)^2} +  (\D \mu _i)_M + (\D \mu _i)_V,
                  \;\;  ~~~~ i = u, d, s.
\label{eq:mu}
\end{equation}
where $(\D \mu _i)_M $ is the contribution from $\va _k$ 
(eq.\ref{eq:kin}), which is evaluated straightforwardly: 
\begin {equation}
(\D \mu _i)_M \equiv 
\f{\p \va_k }{\p M_i}\f{\p  M_i}{\p \rho_i }
=  (\rho _s^u +\rho _s ^d +\rho_s^s)\f{\p M_i}{\p \rho _i}
\end {equation}
$\rho_s^i$ being the scalar density for the {\it i-th} quark 
\begin{equation}
\rho _s ^i = 
\f{3} {4 \pi ^2 } \Big[ 2 M_i k^f_i \sqrt{(k^f_i)^2 + M_i^2} - 2
M_i^3 ln \f{\sqrt{(k^f_i)^2+M_i^2}+k^f_i}{M_i} \Big]
\label{eq:scden}
\end{equation}

On the other hand, the contribution from the potential part (\ref{eq:PE})
of the energy density 
\begin {equation}
(\D \mu _i)_V = \f{\p \va_v }{\p k^f_i}\f{\p k^f_i}{\p \rho _i}+
\f{\p \va_v }{\p M_i}\f{\p  M_i}{\p \rho_i }
\label{eq:delmu}
\end {equation}
is rather complicated and it is evaluated numerically. 
Equations (\ref{eq:sd},\ref{eq:ch}) can now be satisfied and the EOS for 
$\beta$--stable SM is obtained.  

With this EOS we solve the Tolman-Oppenheimer-Volkov equation to 
calculate the structure  of non-rotating SS. 
The properties of the maximum mass configuration for different choices 
of our model parameters are summarized in Tab.1. 
Our EOS is most sensitive to the parameters $\nu$ and $\al _0$ which 
rule the density dependence of the quark mass and DS length respectively. 
However, we found that a change within 20\% of these parameters, 
{\it e.g.}, around the values $\nu = 0.33$, $\al _0 = 0.20$,  produce a 
change of $M_{max}$ and of the corresponding radius which is smaller 
than 10\%.  In the same table we also report, for comparison, the 
results we obtained using an EOS for SM based on the bag model with 
the following choice of the parameters: $B = 110\;MeV/fm^3$, 
$m_s = 150\;MeV$, and $B = 110\;MeV/fm^3$, $m_s = 0$, with 
the strong coupling constant $\alpha_c = 0$ (no gluons) in both cases. 
There have been other theoretical approaches to the study of SS
\cite{dth,bcp}.  The work of Drago et al. \cite{dth} uses ({\it u,d}) 
quark masses close to 100 MeV in the colour dielectric model and gets a  
maximum mass $\sim\;1.6\;M/M_{\odot}$ and a radius of about 10 km.  
In \cite{bcp} a very strong magnetic field is proposed as a mechanism for 
softening a bag equation of state.

%%%%%%%%%%%%%%%%%%%%%%%%%%%%%%%%%%%%%%%%%%%%%%%%%%%%%%%%%%%%%%%%%%%
%%%%%%%%%%%%%%%%%%    Figure 2
%%%%%%%%%%%%%%%%%%%%%%%%%%%%%%%%%%%%%%%%%%%%%%%%%%%%%%%%%%%%%%%%%%%
%%%%%%%%%%%%%%%%%%%%%%%%%%%%%%%%%%%%%%%%%%%%%%%%%%%%%%%%%%%%%%%%%%%

   The calculated  mass-radius relations are plotted in Fig.2, 
and refer to our present model (curves labeled eos1 and eos2) 
and to the bag model for the EOS.  
There is qualitative agreement with the bag model results, and in 
particular
for low values of the mass, $M$ is proportional to $R^3$.  
However, the two bag model calculations, close to our curves indeed use 
a large bag pressure {\it which remains constant at all densities}. 
In Fig.2, we also compare the theoretical MR relations with the 
semiempirical MR for the two strange star candidates. 
The closed region in Fig.2, labeled 4U~1820-30 represents the 
semiempirical MR relation recently extracted from observational data by 
Haberl and Titarchuk \cite{ht} and used in the theoretical analysis of 
ref.\cite{b}.  
The trapezium--like  region labeled Her X-1 represents the semiempirical  
MR relation for the compact object in Her X-1.  
We followed the analysis by Wasserman and Shapiro \cite{ws} updated with 
new mass  and distance measures of Her X-1 as reported in 
ref.\cite{reyn,ker}. 
Dashed curves $a$ and $b$  in Fig.2  denote the MR relation for Her X-1, 
assuming an x-ray luminosity  $L = 2.0\cdot10^{37}$ erg/s and  
$L = 5.0\cdot10^{37}$ erg/s respectively.  
The two values of the luminosity we used correspond to a distance of 
5.0 kpc and 7.9 kpc respectively. 
To be aware of the sensitivity to the luminosity, we also report the MR 
relation (curve $c$) assuming  the Eddington luminosity for 
spherical accretion as an upper bound. 
The two horizontal lines locate the measured mass of Her X-1 
(1.1 -- 1.8) $M_{\odot}$.  

  It is very important to stress that the above MR relations  have been 
extracted from two different type of astronomical phenomena --- x-ray 
burst spectra  (4U~1820-30) and cyclotron line data from a x-ray pulsar 
in a binary system (Her X-1) ---  and using different theoretical models 
to analyze the original observational data. 
The  two semiempirical MR relations overlap in a region of the MR plane 
indicating the existence of a compact object with a radius of 6--8 km. 
This shows that the analysis performed in ref. \cite{b} in the case 
of the x-ray buster 4U~1820-30 also extends  to the case of Her X-1.  
In particular, neutron star models based on ``conventional'' EOS of 
dense matter \cite{b} are unable to reproduce the semiempirical MR 
relation  for these two compact objects. 

  The MR relation calculated with our EOS for SM is well within both 
the semiempirical MR relations of the two SS candidates. 
In conclusion, our present results confirm and reinforce the recent 
claim\cite{li,b} that the compact objects associated with the x-ray 
pulsar Her X-1, and with the x-ray burster 4U~1820-30 are strange stars.  

%%%%%%%%%%%%%%%%%%%%%%%
\vskip 0.5cm
%%%%%%%%%%%%%%%%%%%%%%%
This paper was inspired by  discussions of an early work done by two of 
us (JD and MD) with Prof. Jean Le Tourneux in 1984 (\cite{ddt}). 
JD and MD acknowledge the hospitality of the International Centre 
of Theoretical Physics, Trieste, Italy where this work was inititated.  
MD is particularly
grateful to Dr. M. Malheiro for the introduction of the subject through
series of e-mails, specially, the units and the TOV equation.  It is a
pleasure to thank Drs. S. K. Samaddar, A. Ganguly, B. Datta and F. Weber, S.Raha and
S. Ghosh 
for many discussions.

%%%%%%%%%%%%%%%%%%%%%%%%%%%%%%%%%%%%%%%%%%%%%%%%%%%%%%%%%%%%%%%%%%%%%
%%%                REFERENCES
%%%%%%%%%%%%%%%%%%%%%%%%%%%%%%%%%%%%%%%%%%%%%%%%%%%%%%%%%%%%%%%%%%%%%%%%%%

%%%%%%%%%%%%%%%%%%%%%%%%%%%%%%%%%%%%%%%%%%%%%%%%%%%%%%%%%%%%%%%%%%
%%%%%%%%%%%%%%%%%%%%%%%%%%%%%%%%%%%%%%%%%%%%%%%%%%%%%%%%%%%%%%%%%%%%%%%%%%
%%%%%%%%%%%%%%%%%%%%%%%%%%%%%%%%%%%%%%%%%%%%%%%%%%%%%%%%%%%%%%%%%%%%%%%%%%
 \newpage
%%%%%%%%%%%%%%%%%%%%%%%%%%%%%%%%%%%%%%%%%%%%%%%%%%%%%%%%%%%%%%%%%%%%%%%%
%%%%%%%%%%%%%%%%%     Table 1
%%%%%%%%%%%%%%%%%%%%%%%%%%%%%%%%%%%%%%%%%%%%%%%%%%%%%%%%%%%%%%%%%%%%%%%%
\begin{table}
\caption{Properties of  the maximum mass strange star configuration obtained
for different equations of state: 
$M_G$ is the gravitational (maximum) mass, R is the corresponding radius,  
$\rho_c$ the central number density, $\varepsilon_c$ the central mass density. 
Our EOS for different choices of the parameters are denoted as follow: 
(eos1) $\nu = 0.40$, $\al _0 = 0.20$;
(eos2) $\nu = 1/3$,  $\al _0 = 0.20$; 
(eos3) $\nu = 1/3$,  $\al _0 = 0.15$. 
The the bag model EOS are denoted as follow: 
(B110;150) $B = 110\;MeV/fm^3$, $m_s = 150\;MeV$; 
(B110;0)  $B = 110\;MeV/fm^3$, $m_s = 0$  ($\alpha_c = 0$ in both cases) }
\vskip 1cm
\begin{center}
\begin{tabular}{|c|c|c|c|c|}
\hline\\
EOS & $M_G/M_{\odot}$ & R(km) & $\rho_c(fm^{-3})$  & 
$\varepsilon_c( 10^{14}g/cm^{3})$\\ 
\hline\\
 eos1            & 1.414 & 7.18 & 2.24 &  44.6 \\
 eos2            & 1.286 & 6.53 & 2.57 &  53.6 \\
 eos3            & 1.301 & 6.62 & 2.55 &  52.4 \\
 B110;150        & 1.358 & 7.50 & 1.87 &  42.0 \\
 B110;0          & 1.448 & 7.90 & 1.76 &  38.1 \\
\hline 
\end{tabular}
\end{center}
\end{table}
%%%%%%%%%%%%%%%%%%%%%%%%%%%%%%%%%%%%%%%%%%%%%%%%%%%%%%%%%%%%%%%%%%%%%%%%

%%%%%%%%%%%%%%%%%%%%%%%%%%%%%%%%%%%%%%%%%%%%%%%%%%%%%%
%%%%%%%\end{multicols}
%%%%%%%%%%%%%%%%%%%%%%%%%%%%%%%%%%%%%%%%%%%%%%%%%%%%%%

%%%%%%%%%%%%%%%%%%%%%%%%%%%%%%%%%%%%%%%%%%%%%%%%%%%%%%%%%%%%%%%%%%%%%%%%%%
  \newpage
%%%%%%%%%%%%%%%%%%%%%%%%%%%%%%%%%%%%%%%%%%%%%%%%%%%%%%%%%%%%%%%%%%%%%%%%%%
%        FIGURE CAPTIONS
%%%%%%%%%%%%%%%%%%%%%%%%%%%%%%%%%%%%%%%%%%%%%%%%%%%%%%%%%%%%%%%%%%%%%%%%%%
\vskip 0.3cm
{\bf Fig. 1} The inverse Debye screening length $D^{-1}$ for 
$\alpha_0 = 0.20$, and the $d$ quark mass $M_d$, for $\nu = 0.4$ (dashed line) 
and $\nu = 1/3$ (continuous line),  are plotted as a function of the baryon 
number density.  
\vskip 0.3cm
{\bf Fig. 2} The theoretical MR relations (curves eos1 and eos2) calculated 
within the present model are compared with the semiempirical MR relations 
for 4U~1820-30 and Her~X-1 (closed regions in the MR plane labeled  
4U~1820-30 and Her~X-1, respectively).  The remaining two continuous curves 
represent the MR relation calculated with the bag model EOS, with 
$m_s = 0$ (upper curve) and $m_s = 150$ MeV (lower curve), 
and $B = 110\;MeV/fm^3$ in both cases. 
\vskip 0.4cm
%%%%%%%%%%%%%%%%%%%%%%%%%%%%%%%%%%%%%%%%%%%%%%%%%%%%%%%%%%%%%%%%%%%%%%%%%%
%%%%%%%%%%%%%%%%%%%%%%%%%%%%%%%%%%%%%%%%%%%%%%%%%%%%%%
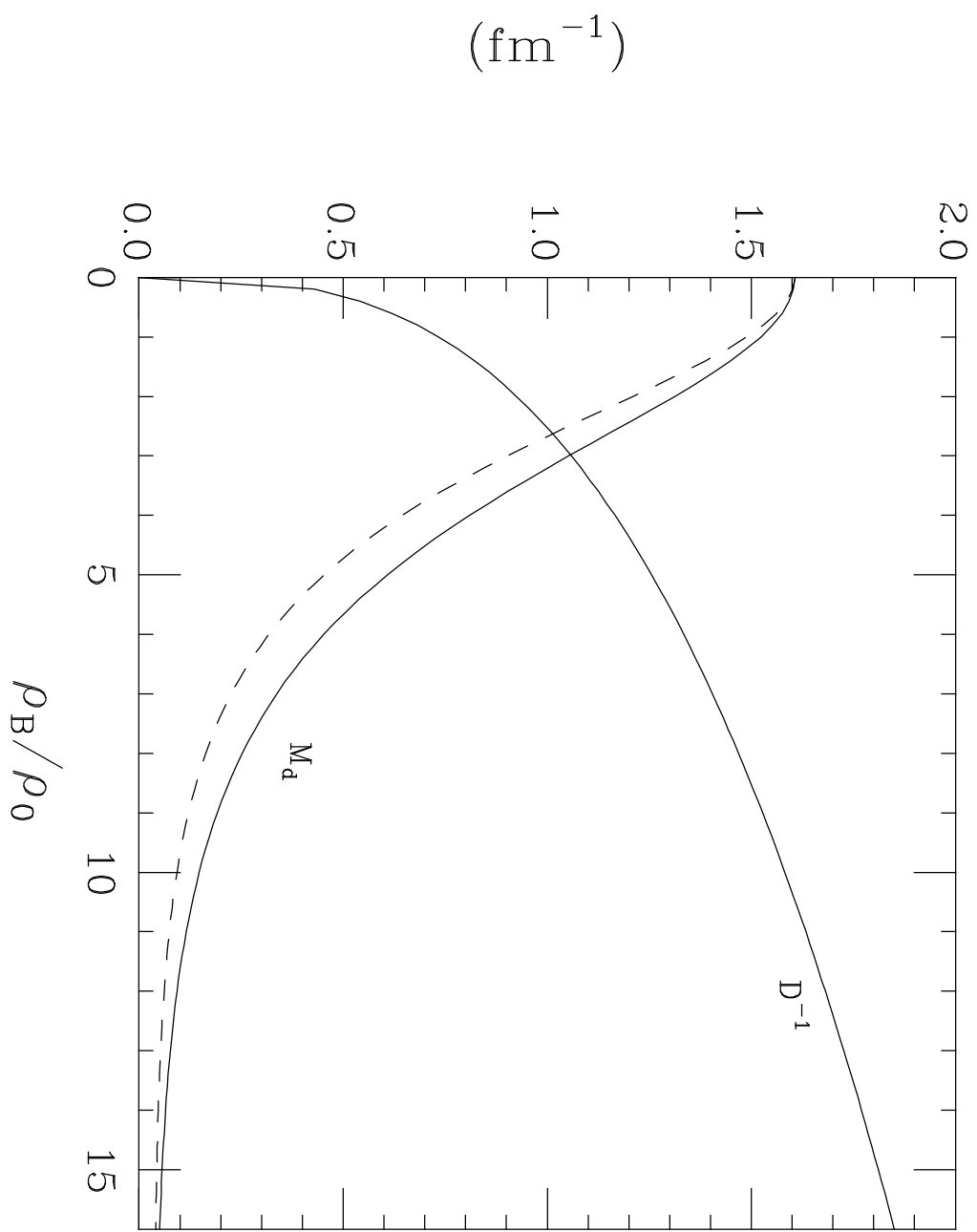
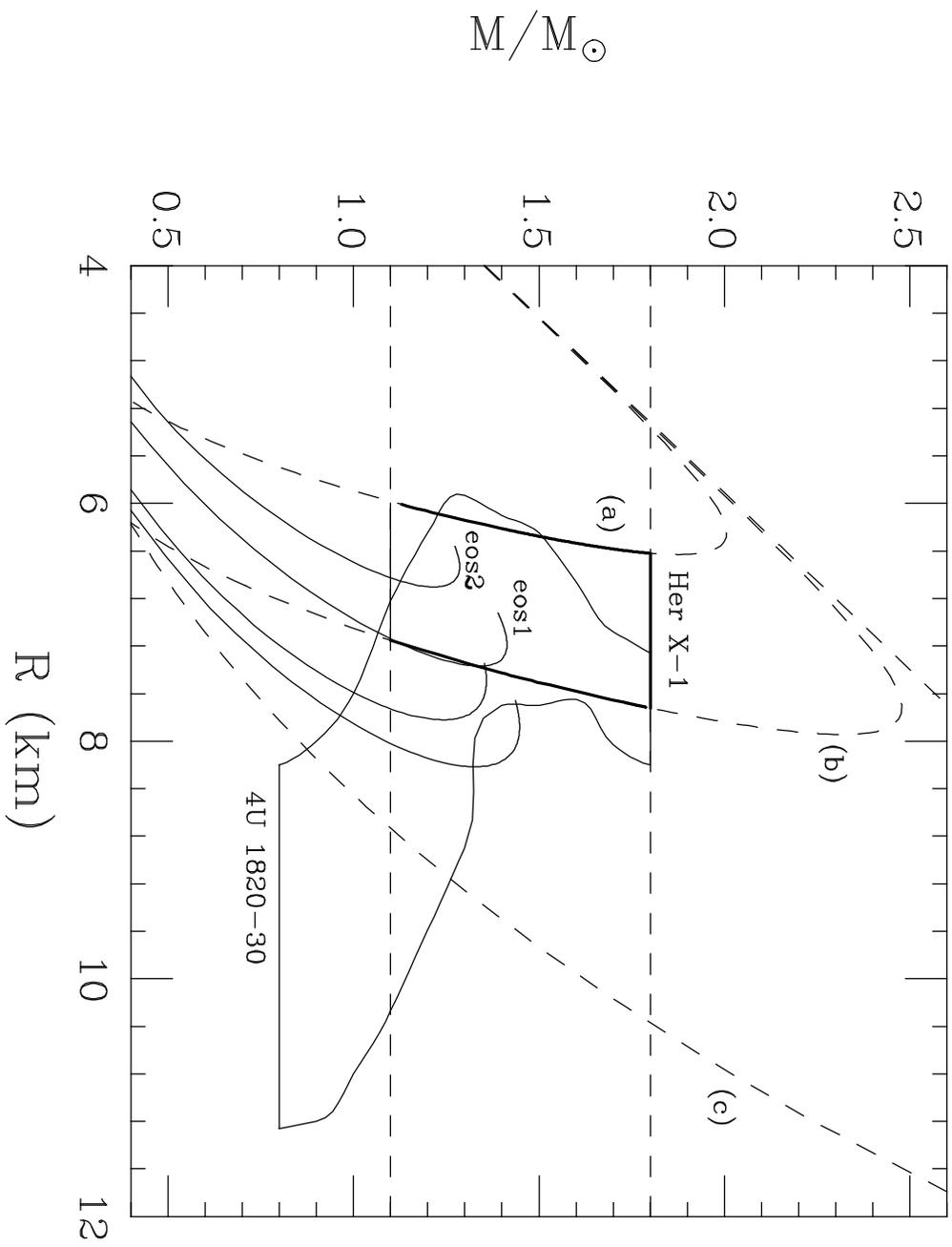
\end{document}